\documentclass[conference]{IEEEtran}
\IEEEoverridecommandlockouts
\usepackage{cite}
\usepackage{amsmath,amssymb,amsfonts}
\usepackage{epsfig}
\usepackage{theorem}
\usepackage{algorithm}
\usepackage{algorithmic}
\usepackage{graphicx}
\usepackage{textcomp}
\usepackage{xcolor}
\usepackage{mathtools}

\def\BibTeX{{\rm B\kern-.05em{\sc i\kern-.025em b}\kern-.08em
    T\kern-.1667em\lower.7ex\hbox{E}\kern-.125emX}}

\def \C {\mathbb C}

\def \R {\mathbb R}

\DeclareMathOperator{\rect}{rect}

\newcommand{\wML}{\mathbf{w}_{\text{ML}}}
\newcommand{\wSL}{\mathbf{w}_{\text{SL}}}
\newcommand{\herm}{^{\text{H}}}
\newcommand{\trans}{^{\text{T}}}
\newcommand{\fracpar}[2]{\frac{\partial #1}{\partial #2}}
\newcommand{\lb}{{\ell}}

\newcommand\nth{^{\text{th}}}
\newcommand\bphi{\boldsymbol\phi}
\newcommand{\num}{\wSL\trans\vert \mathbf{r} \vert^p}
\newcommand{\den}{\wML\trans \vert \mathbf{r} \vert^p}
\newcommand{\grad}[2]{\nabla_{#1}{#2}}

\begin{document}

\title{Gradient-Descent Based Optimization of Constant Envelope OFDM Waveforms
\thanks{David G. Felton's efforts were supported by the Naval Research Enterprise Internship Program (NREIP) and David A. Hague's efforts were by the Naval Undersea Warfare Center's In-House Laboratory Independent Research (ILIR) program.}
}

\author{\IEEEauthorblockN{David G. Felton$^1$ David A. Hague$^2$}
\IEEEauthorblockA{\textit{$^1$Radar Systems Lab (RSL), University of Kansas, Lawrence, KS}\\
\textit{$^2$Sensors and Sonar Systems Department, Naval Undersea Warfare Center, Newport, RI}}
}
\maketitle

\begin{abstract}
This paper describes a gradient-descent based optimization algorithm for synthesizing Constant Envelope Orthogonal Frequency Division Multiplexing (CE-OFDM) waveforms with low Auto-Correlation Function (ACF) sidelobes in a specified region of time-delays.  The algorithm optimizes the Generalized Integrated Sidelobe Level (GISL) which controls the mainlobe and sidelobe structure of the waveform's ACF.  The operations of this Gradient-Descent GISL (GD-GISL) algorithm are FFT-based making it computationally efficient.  This computational efficiency facilitates the design of large dimensional waveform design problems.  Simulations demonstrate the GD-GISL algorithm on CE-OFDM waveforms employing Phase-Shift Keying (PSK) symbols that take on a continuum of values (i.e, $M_{\text{PSK}} = \infty$).  Results from these simulations show that the GD-GISL algorithm can indeed reduce ACF sidelobes in a desired region of time-delays.  However, truncating the symbols to finite M-ary alphabets introduces perturbations to the waveform's instantaneous phase which increases the waveform's ACF sidelobe levels.  
\end{abstract}


\section{Introduction}
The Constant-Envelope Orthogonal Frequency Division Multiplexing (CE-OFDM) waveform is a constant envelope analogue of the standard OFDM waveform model.  The OFDM modulation is performed in either the instantaneous phase or frequency domain \cite{Thompson_2, Chung_Comm} and is essentially a form of Frequency Modulation (FM) which guarantees a constant envelope.  This makes the waveform better suited for transmission on real-world radar and communications transmitters than standard OFDM waveforms \cite{Thompson_1} whose complex envelope can vary substantially \cite{Levanon}.  CE-OFDM waveforms have recently been proposed for use in Dual-Function Radar/Communication (DFRC) applications \cite{Aubry_2016_Optimization, IEEE_Sig_Proc_RadarComm, Zhang_Radar_Comms} as the encoded communication symbols can also be utilized as a discrete set of parameters that can realize optimized constant modulus radar waveforms \cite{Gini_2012_Waveform, Rigling}.  

Recent efforts in the literature have studied CE-OFDM as a potential radar waveform and analyzed the structure of the waveform's Ambiguity Function (AF) and Auto/Cross-Correlation Functions (ACF/CCF) \cite{Mohseni, Blunt_CEOFDM}.  From a waveform design perspective, the OFDM symbols represent a discrete set of parameters that can be optimized to synthesize CE-OFDM waveforms with desirable AF/ACF characteristics.  However, to the best of the authors' knowledge, optimization methods to synthesize novel CE-OFDM waveform designs have not yet been developed.  This paper introduces a gradient-descent based algorithm that synthesizes CE-OFDM waveforms with low ACF sidelobes in a specified region of time-delays via minimization of a Generalized Integrated Sidelobe Level (GISL) metric.  This Gradient-Descent GISL (GD-GISL) algorithm leverages methods developed in \cite{Blunt_PCFM_Grad} that were used to optimize Polyphase-Coded FM (PCFM) waveforms.   Since the GD-GISL algorithm's operations are largely comprised of FFTs, it is computationally efficient which facilitates synthesizing large-dimensional waveform design problems.  

The capabilities of this GD-GISL algorithm to optimize CE-OFDM waveforms are demonstrated by several illustrative design examples.  These examples use CE-OFDM waveforms that employ Phase-Shift Keying (PSK) where the symbols reside on the unit-circle and take on a continuum of values (i.e, $M_{\text{PSK}} = \infty$).  The results from these design examples show that the CE-OFDM waveform model can indeed be optimized to possess lower ACF sidelobes in a specific region of time-delays than pseudo-randomly generated PSK coefficients.  However, truncating the CE-OFDM symbols to finite M-ary alphabets will increase the waveform's ACF sidelobes.  The rest of this paper is organized as follows: Section II descibes the CE-OFDM waveform model and the design metrics used to assess the waveform's ACF characteristics.  Section III describes the GD-GISL algorithm.  Section IV evaluates the performance of this algorithm via several illustrative design examples.  Finally, Section V concludes the paper. 

\section{CE-OFDM Waveform Model}
A basebanded FM waveform with unit energy is expressed as
\begin{equation} \label{eq:sig_mod}
s\left(t\right)=\frac{\rect\left(t/T\right)}{\sqrt{T}} e^{j\varphi(t)} 
\end{equation}
where $\varphi\left(t\right)$ is the waveform's phase modulation function, $T$ is the waveform's duration, and $1/\sqrt{T}$ normalizes the signal energy to unity.  A CE-OFDM waveform utilizing Phase-Shift Keying (PSK) possesses a phase modulation function that is expressed as \cite{Blunt_CEOFDM, Hague_Felton_CE_OFDM_1}
\begin{equation}
\varphi(t)=2\pi h\sum_{\ell=1}^L{\vert\Gamma_\lb\vert\cos\left(\dfrac{2\pi \ell t}{T}+\phi_\lb\right)}
\label{eq:ceofdmPhase}
\end{equation}
where $L$ is the number of phase modulation sub-carriers, $\vert\Gamma_\lb \vert= 1$ and $\phi_\lb$ are the magnitude and phase respectively of the CE-OFDM symbol associated with the $\ell^{\text{th}}$ subcarrier, and $h$ is the modulation index which along with $L$ determines the waveform's  bandwidth \cite{Hague_Felton_CE_OFDM_1}.  The waveform's corresponding frequency modulation function $m\left(t\right)$ is expressed as
\begin{equation}
m\left(t\right)=\frac{1}{2\pi}\frac{\partial\varphi\left(t\right)}{\partial t}=-\dfrac{2\pi h}{T}\sum_{\ell=1}^L \ell\vert\Gamma_\lb\vert\sin\left(\dfrac{2\pi \ell t}{T}+\phi_\lb\right).
\label{eq:modFunc}
\end{equation}
The CE-OFDM symbols $\phi_\lb$ are utilized as a discrete set of design parameters.  Modifying these parameters in an intelligent manner can produce waveforms with low AF/ACF sidelobes.  Additionally, the CE-OFDM's phase and frequency modulation functions are expressed as a finite Fourier series which are infinitely differentiable \cite{boyd}.  This property makes these functions smooth and devoid of any transient components resulting in the vast majority of the CE-OFDM waveform's energy being densely concentrated in a compact band of frequencies.  Coupling this spectral compactness property with its natural constant envelope makes the CE-OFDM waveform model well-suited for transmission on real-world radar transmitter devices.  The design versatility and the transmitter amenability of the CE-OFDM waveform model makes it an attractive choice for a variety of radar applications.  

This paper assumes a Matched-Filter (MF) receiver is used to process the target's echo signal.  The narrowband AF measures the MF response to Doppler shifted versions of the transmit waveform and is expressed as
\begin{equation}
\chi\left(\tau, \nu\right) = \int_{-\infty}^{\infty} s\left(t-\frac{\tau}{2}\right)s^*\left(t+\frac{\tau}{2}\right) e^{j2\pi \nu t}dt
\label{eq:AF}
\end{equation}
where $\nu$ is the Doppler shift.  The zero-Doppler cut of the AF, the ACF, provides the range response of the waveform's MF output and is expressed as
\begin{equation}
R\left(\tau\right) = \chi\left(\tau, \nu\right)|_{\nu=0} = \int_{-\infty}^{\infty} s\left(t-\frac{\tau}{2}\right)s^*\left(t+\frac{\tau}{2}\right) dt.
\label{eq:ACF}
\end{equation}

There are several metrics that describe the sidelobe structure of a waveform's ACF.  Two of the most common metrics are the Peak-to-Sidelobe Level Ratio (PSLR) and the Integrated Sidelobe Level (ISL).  The PSLR is expressed as
\begin{IEEEeqnarray}{rCl}
\text{PSLR} =  \dfrac{\underset{\Delta \tau \leq \vert\tau\vert \leq T}{\text{max}}\bigl\{\left|R\left(\tau\right)\right|^2\bigr\}}{\underset{0 \leq \vert\tau\vert \leq \Delta \tau}{\text{max}}\bigl\{\left|R\left(\tau\right)\right|^2\bigr\}} = \underset{\Delta \tau \leq \vert\tau\vert \leq T}{\text{max}}\bigl\{\left|R\left(\tau\right)\right|^2\bigr\}
\label{eq:PSLR}
\end{IEEEeqnarray} 
where $\Delta_{\tau}$ denotes the first null of the ACF and therefore establishes the mainlobe width of the ACF as $2\Delta\tau$.  Note that the rightmost expression in \eqref{eq:PSLR} results from the assumption that the waveform is unit energy and thus the maximum value of $|R\left(\tau\right)|^2$ is unity which occurs at $\tau = 0$.  The ISL is the ratio of the area under the sidelobe region $A_{\tau}$ of $|R\left(\tau\right)|^2$ to the area under mainlobe region $A_0$ of $|R\left(\tau\right)|^2$ expressed as
\begin{IEEEeqnarray}{rCl}
\text{ISL}~=\dfrac{A_{\tau}}{A_0} = \dfrac{\int_{\Omega_{\tau}}\left|R\left(\tau\right)\right|^2 d\tau}{\int_{-\Delta \tau}^{\Delta \tau}\left|R\left(\tau\right)\right|^2 d\tau}
\label{eq:ISL}
\end{IEEEeqnarray}
where the term $\Omega_{\tau}$ denotes any sub-region of time-delays of the ACF excluding the mainlobe region $-\Delta\tau \leq \tau \leq \Delta \tau$.  A lower ISL corresponds to an ACF with lower overall sidelobe levels in the region $\Omega_{\tau}$ and/or a larger area $A_0$ under the mainlobe region which implies that the mainlobe width $2\Delta\tau$ increases.   Note that while reducing the area $A_{\tau}$ will generally reduce sidelobe levels, it does not always directly translate to a lower PSLR \cite{Blunt_Waveform_Diversity}.

The Generalized Integrated Sidelobe Level (GISL) \cite{Blunt_PCFM_Grad} generalizes the ISL metric in \eqref{eq:ISL} by evaluating the $\ell_p$-norm \cite{PrabhuBabuI, PrabhuBabuIV} of the sidelobe and mainlobe regions of the ACF expressed as
\begin{IEEEeqnarray}{rCl}
\text{GISL}~= \left(\dfrac{\int_{\Omega_{\tau}}\left|R\left(\tau\right)\right|^p d\tau}{\int_{-\Delta \tau}^{\Delta \tau}\left|R\left(\tau\right)\right|^p d\tau}\right)^{2/p}
\label{eq:GISL}
\end{IEEEeqnarray} 
where $p \geq 2$ is an integer.  When $p=2$, the GISL becomes the standard ISL metric.  As $p\rightarrow\infty$, the integrals in \eqref{eq:GISL} approach the infinity norm $||\cdot||_{\infty}^2$, also known as the max-norm, and correspondingly the GISL approaches the PSLR metric \eqref{eq:PSLR} \cite{Blunt_PCFM_Grad}.  However, from a waveform optimization perspective, the max-norm can produce a discontinuous objective function which prevents the efficient use of gradient-descent based waveform optimization methods.  Making $p$ large but finite results in a smooth objective function which facilitates the use of gradient-descent based methods in the waveform optimization problem.  Empirical analysis \cite{Blunt_PCFM_Grad} suggests that values of $p \geq 6$ or so produces such a PSLR-like metric.  For this reason the GISL is the design metric this paper uses to optimize the ACF sidelobe levels of CE-OFDM waveforms.

\begin{figure}[ht]
\centering
\includegraphics[width=0.5\textwidth]{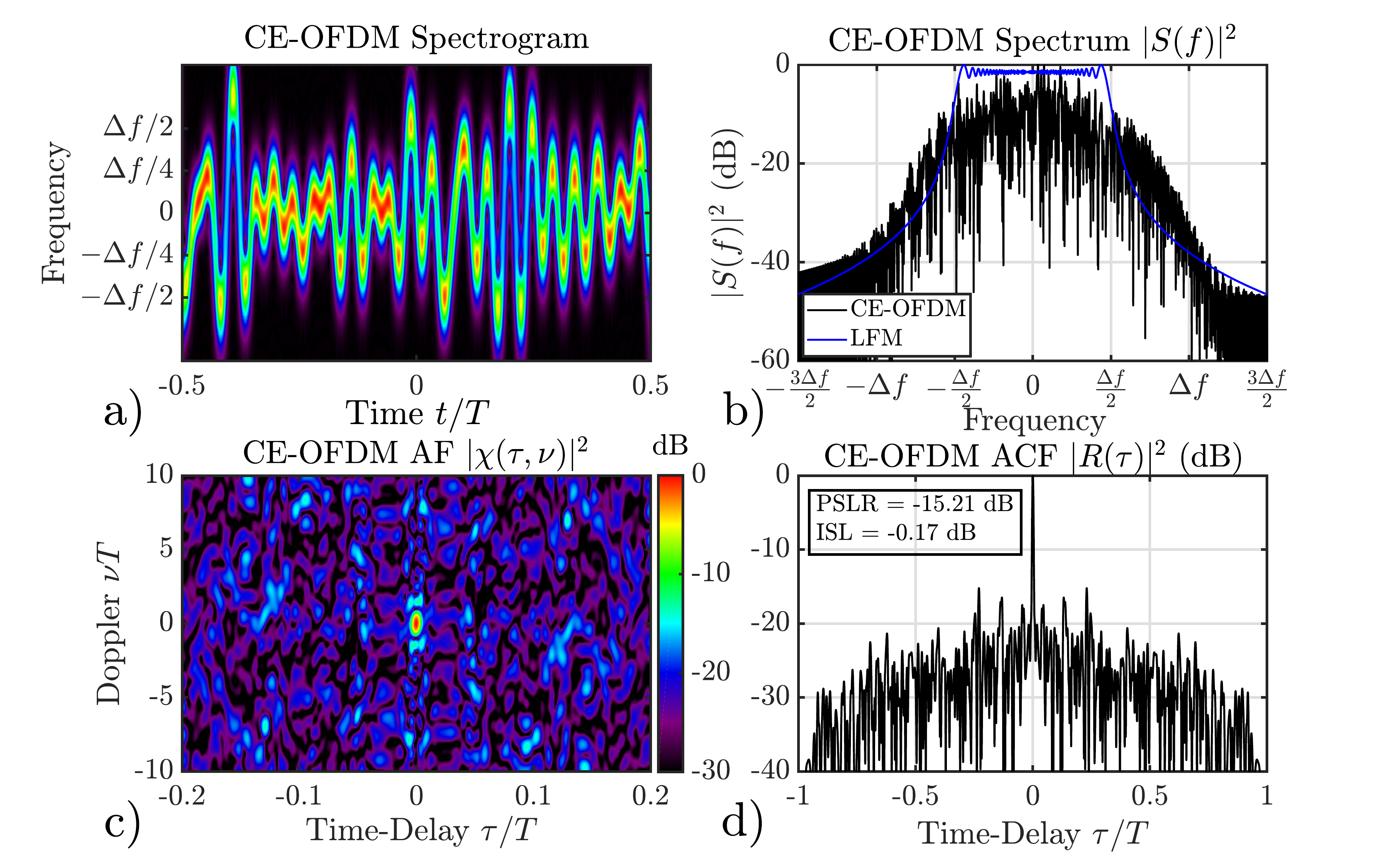}
\caption{Spectrogram (a), spectrum (b), AF (c), and ACF (d) of a CE-OFDM waveform with a TBP of 200 ($h= 0.1856$) and $L=24$ randomly generated PSK coefficients where $M_{\text{PSK}}=32$.  Also shown in panel (b) is the spectrum of a LFM waveform of the same TBP.  The CE-OFDM waveform possesses a spectrum that is densely concentrated in a compact band of frequencies and possesses a ``Thumbtack-Like'' AF shape.}
\label{fig:CEOFDM_1}
\end{figure}

Figure \ref{fig:CEOFDM_1} shows an example CE-OFDM waveform's spectrogram, spectrum, AF, and ACF.  The waveform is composed of $L=24$ sub-carriers that employ PSK symbols $\phi_\lb$ with $M_{\text{PSK}}=32$ generated from a pseudo-random sequence as described in \cite{Blunt_CEOFDM}.  The modulation index $h=0.1856$ is chosen to synthesize a waveform with the same RMS bandwidth \cite{Cook} as that of a LFM waveform with a Time-Bandwidth Product (TBP) of 200 which can be calculated in closed form as \cite{Hague_Felton_CE_OFDM_1}
\begin{equation}
h = \frac{T\Delta f}{2\pi \sqrt{2L^3+3L^2+L}}.
\end{equation}
As explained earlier, the CE-OFDM waveform's frequency modulation function is a finite Fourier series and is therefore an infinitely differentiable and smooth function.  This smooth FM function is shown in panel (a) of Figure \ref{fig:CEOFDM_1}.  As a result of the smooth FM function, the vast majority of the waveform's energy is densely concentrated in a compact band of frequencies as is shown in panel (b) of Figure \ref{fig:CEOFDM_1}.  As is described in \cite{Hague_Felton_CE_OFDM_1}, a CE-OFDM waveform utilizing PSK coding will essentially always produce a ``Thumbtack-Like'' AF shape.  The sidelobe pedestal is clearly visible in the ACF of the waveform shown in panel (d).  It is the goal of this paper to develope an algorithm that will further reduce the ACF sidelobe levels in a specified region $\Omega_{\tau}$ of time-delays.  

\newpage

\section{The GD-GISL Algorithm}
\label{sec:GISL}
The first step in developing the GD-GISL algorithm is to discretize the waveform signal model and its design metrics.  The CE-OFDM waveform's instantaneous phase \eqref{eq:ceofdmPhase} can be transformed from the amplitude-phase Fourier series representation to the standard real-valued Fourier series as
\begin{equation}
\varphi\left(t\right) = 2\pi h \sum_{\ell=1}^L \tilde{\alpha}_{\ell}\cos\left(\dfrac{2\pi \ell t}{T}\right)+\tilde{\beta}_{\ell}\sin\left(\dfrac{2\pi \ell t}{T}\right).
\label{eq:ceofdm_1}
\end{equation}
The Fourier coefficients $\tilde{\alpha}_{\ell}$ and $\tilde{\beta}_{\ell}$ lie on the unit circle such that $\vert\Gamma_{\ell}\vert = \sqrt{\tilde{\alpha}_{\ell}^2 + \tilde{\beta}_{\ell}^2} = 1$ and are derived from $\phi_{\ell}$
\begin{align}
\tilde{\alpha}_{\ell} &= \vert\Gamma_{\ell}\vert \cos\phi_{\ell}, \\
\tilde{\beta}_{\ell} &= \vert\Gamma_{\ell}\vert \sin\phi_{\ell}.
\end{align}
From here, \eqref{eq:ceofdm_1} can be written as a linear sum using discrete variables as
\begin{equation}
	\mathbf{\varphi} = \begin{bmatrix}
		\mathbf{B}_\mathrm{c} & \mathbf{B}_\mathrm{s}
	\end{bmatrix} \begin{bmatrix}
	\cos\left(\boldsymbol\phi\right) \\ \sin\left(\boldsymbol\phi\right)
\end{bmatrix} 
\end{equation} 
where the phase values are grouped into a $L\times1$ vector $\boldsymbol\phi= \left[\phi_1, \phi_2, \dots, \phi_L \right]\trans$ and the $ M{\times}L $ basis matrices $\mathbf{B}_\mathrm{c}$ and $\mathbf{B}_\mathrm{s}$  contain cosine and sine harmonics respectively such that the $ \ell^{\text{th}} $ columns
\begin{align}
\mathbf{b}_{\mathrm{c}, \lb} &= \cos\left(\dfrac{2\pi \lb t}{T} \right), \\
\mathbf{b}_{\mathrm{s}, \lb} &= \sin\left(\dfrac{2\pi \lb t}{T} \right)
\end{align}
are sampled at some sampling rate $f_s$ which satisfies the Nyquist criterion. 

From here, the development of the gradient-based GISL algorithm largely follows the description given in \cite{Blunt_PCFM_Grad}.  The GISL metric can be expressed in terms of the discretized ACF which is expressed as
\begin{equation} \label{eq:discr}
\mathbf{r} = \mathbf{A}\herm \lvert \mathbf{A\bar{s}} \rvert^2
\end{equation}
where $\mathbf{r} \in \C^{(2M-1)}$  contains discretized samples of the ACF, $\mathbf{\bar{s}} \in \C^{(2M-1)}$ is a discretized and zero-padded version of $\mathbf{s}$, and $\mathbf{A}$ and $\mathbf{A}\herm$ are $(2M-1) \times (2M-1)$ Discrete Fourier Transform (DFT) and Inverse DFT matrices respectively. The GISL metric is then expressed as the cost function \cite{Blunt_PCFM_Grad}
\begin{equation}
J_p = \dfrac{\|\wSL \odot \mathbf{r} \|_p^2}{\|\wML \odot \mathbf{r} \|_p^2}
\label{eq:GISLmyNISL}
\end{equation}
where the vectors $\wSL$ and $\wML$ $\in \R^{(2M-1)}$ are non-zero in the extent of the sidelobe and mainlobe regions respectively.  In setting the cost function $J_p$ to the GISL metric, the optimization problem can be formally stated as 
\begin{equation}
\underset{\bphi}{\text{min}}~J_p.
\label{eq:Problem_1}
\end{equation}

The GISL is an L-dimensional and highly non-convex objective function across the CE-OFDM parameter space $\phi_\lb$ on the order of $2p$ in each dimension.  Therefore, convergence to the global minimum is most certainly not guaranteed.  Non-convex functions require a non-linear optimization routine, which is why here we expand upon the continuous nature of the CE-OFDM waveform and its modulation function to employ gradient-descent optimization.  Gradient-descent is an iterative approach which takes some step $\mu$ in the direction of steepest descent $q_i$
\begin{align}
\bphi_{i+1} &= \bphi_i + \mu q_i  \label{eq:phi_i} \\
\mathbf{q}_i &= -\mathbf{\nabla}_{\bphi_i}J_p \label{eq:qi} 
\end{align} 
where $\nabla_{\bphi}$ is the gradient operator.  The gradient of \eqref{eq:Problem_1}, derived in the Appendix, is expressed as
\begin{equation}
	\grad{\bphi}{J_p}=4 J_p \mathbf{\bar{D}}^T \Im{\Biggl\{\mathbf{\bar{s}}^* \odot \mathbf{A}\herm \left[(\mathbf{A\bar{s}}) \odot \mathbf{P} \right] \Biggr\}} 
\label{eq:gradDevil1}
\end{equation}
where 
\begin{equation}
\mathbf{P} = \Re\left\{\mathbf{A} \left( \lvert\mathbf{r}\rvert^{p-2}\odot\mathbf{r}\odot \left[ \frac{\wSL}{\num} - \frac{\wML}{\den} \right]\right)\right\}.
\label{eq:gradDevil2}
\end{equation}
Performing \eqref{eq:phi_i} and \eqref{eq:qi} iteratively until the Euclidean length of $q_i$ is below some threshold $g_{\text{min}}$ ensures that $J_p$ is very near a local minima.  Alternatively, the routine may continue until it reaches a predetermined number of iterations $I_{\text{max}}$.

We employ heavy-ball gradient-descent which includes weighted versions of the previous search-directions with the current gradient.  This has been shown to converge quickly for these types of problems by dampening rapid transitions of the gradient thereby enforcing a smooth path to the minima. The search direction is altered by inclusion of previous gradients as
\begin{equation}
\mathbf{q}_i = -\nabla_{\bphi_i}J_p + \beta \mathbf{q}_{i-1}
\end{equation}
where $\beta \in \left[0, 1\right]$.  Since this method does not always ensure a descent, if in fact the current search direction is an ascent (i.e., the projection of the gradient onto the current search direction is positive), the current search direction is reset to the current gradient.
\begin{equation}
\text{if}~\mathbf{q}_i\trans(\nabla_{\bphi_i}J_p )> 0,~\text{then}~\mathbf{q}_i = -\nabla_{\bphi_i}J_p.
\end{equation}
Once the search direction is established, a simple backtracking method is used to calculate the step size $\mu$ for the line search that satisfies sufficient decrease via the Armijo condition \cite{Armijo}.  As mentioned earlier, to maintain continuity of the GISL gradient and to avoid diminishing returns thereafter, the parameter $p$ should take on values much lower than infinity but larger than roughly 6 \cite{Blunt_PCFM_Grad}.  These values for $p$ have been shown to consistently generate waveforms with a flat sidelobe response.  On the other hand, letting $p=2$ results in waveforms whose ACFs possess more peaks and valleys, but overall less area in the specified sidelobe region $\Omega_{\tau}$.  The steps of the GD-GISL algorithm are listed in Algorithm 1.  Since the algorithm makes extensive use of FFTs in computing the GISL metric \eqref{eq:gradDevil1}, it is substantially more computationally efficient compared to a brute-force numerical computation.  

\begin{algorithm}
 \caption{The GD-GISL Algorithm}
 \begin{algorithmic}[1]
 \renewcommand{\algorithmicrequire}{\textbf{Input:}}
 \renewcommand{\algorithmicensure}{\textbf{Output:}}
 \REQUIRE Initialize $\mathbf{B}$, $\mathbf{\phi}^{(0)}$, $P$, $L$, $\mathbf{q}_0 = \mathbf{0}_{\text{N}\times1}$, $\beta$, $\mu$, $\rho_{\text{up}}$, $\rho_{\text{down}}$, $c$, and set $i=1$.
 \ENSURE  Final CE-OFDM coefficient vector $\bphi$ with refined ACF properties that locally solves the criteria in \eqref{eq:Problem_1}
  \STATE Evaluate $J_p\left(\bphi_{i-1}\right)$ and $\nabla_{\bphi_i}J_p\left(\bphi_{i-1}\right)$ via \eqref{eq:GISLmyNISL} and \eqref{eq:gradDevil1}.
  \STATE $\mathbf{q}_i = -\nabla_{\bphi_i}J_p + \beta \mathbf{q}_{i-1}$
  \STATE \textbf{If} $\left(\nabla_{\bphi_i}J_p\left(\bphi_{i-1}\right) \right)^\text{T}~\mathbf{q}_i \geq 0$
  \STATE ~~~~$\mathbf{q}_i = -\nabla_{\bphi_i}J_p\left(\bphi_{i-1}\right)$
  \STATE \textbf{End}(If)
  \STATE \textbf{While} $ J_p(\bphi_i{+}\mu \mathbf{q}_i  \label{eq:xi}){>}J_p(\bphi_{i-1}) +c\mu\left(\nabla_{\bphi_i}J_p\left(\bphi_{i-1}\right) \right)^\text{T}~\mathbf{q}_i$
  \STATE ~~~~$\mu = \rho_{\text{down}}\mu$
  \STATE \textbf{End}(While)
  \STATE $\bphi_i = \bphi_{i-1} + \mu \mathbf{q}_i,~~\mu = \rho_{\text{up}}\mu$
  \STATE $i = i+1$
  \STATE Repeat steps 1-9 until $i=I$ or $\|\nabla_{\bphi}J_p\left(\bphi_i\right) \| \leq g_{\text{min}}$
 \end{algorithmic} 
 \end{algorithm}

\section{Several Illustrative Design Examples}
\label{sec:Examples}
\begin{figure}[ht]
\centering
\includegraphics[width=0.5\textwidth]{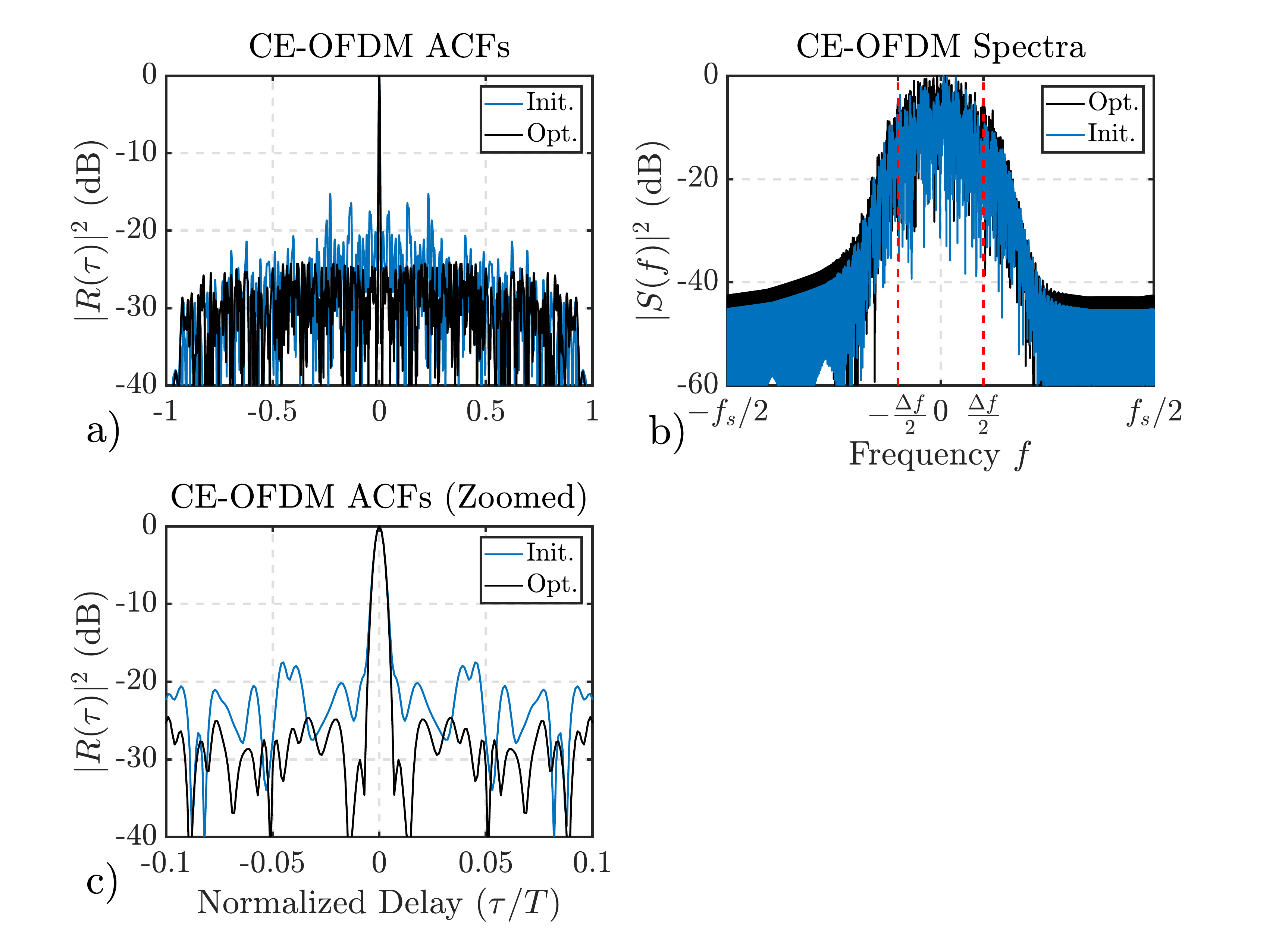}
\caption{ACFs (a), Spectra (b), and zoomed ACFs (c) of initial and optimized CE-OFDM waveforms.  The CE-OFDM waveform resulting from the GISL optimization algorithm possesses clearly lower ACF sidelobes while also maintaining the same mainlobe width as the initial waveform. }
\label{fig:CEOFDM_2}
\end{figure}

This section demonstrates the GD-GISL algorithm using two waveform optimization design examples.  Both examples optimize the GISL ($p=20$) of the waveform from Figure \ref{fig:CEOFDM_1} over different sub-regions $\Omega_{\tau}$ of the waveform's ACF.  The waveform time-series are sampled at a rate $f_s = 5\Delta f$.  The algorithm was set to run for a max number of iterations $I=100$.   Figure \ref{fig:CEOFDM_2} shows the ACF, spectrum, and ACF zoomed in at the origin of the CE-OFDM waveform from Figure \ref{fig:CEOFDM_1} and the resulting optimized CE-OFDM waveform whose ACF sidelobes were minimized over all time-delays $\tau$ excluding the mainlobe region.  The initial waveform's GISL for $p=20$ and PSLR values were -14.73 dB and -15.21 dB respectively.  The optimized waveform's GISL was -22.14 dB corresponding to a PSLR of -20.72 dB.  It's also important to note that the mainlobe width of the optimized waveform stayed the same width as its initial seed waveform.  As explained in \cite{Hague_Felton_CE_OFDM_1}, the CE-OFDM's RMS bandwidth, which controls the ACF mainlobe width, remains the same value for fixed $L$ and $h$ regardless of the PSK values $\phi_{\lb}$.  This precludes introducing constraints on RMS bandwidth to ensure the waveform's ACF mainlobe width stays largely fixed as was done for a closely related waveform model in \cite{Hague_AES}.  

\begin{figure}[ht]
\centering
\includegraphics[width=0.5\textwidth]{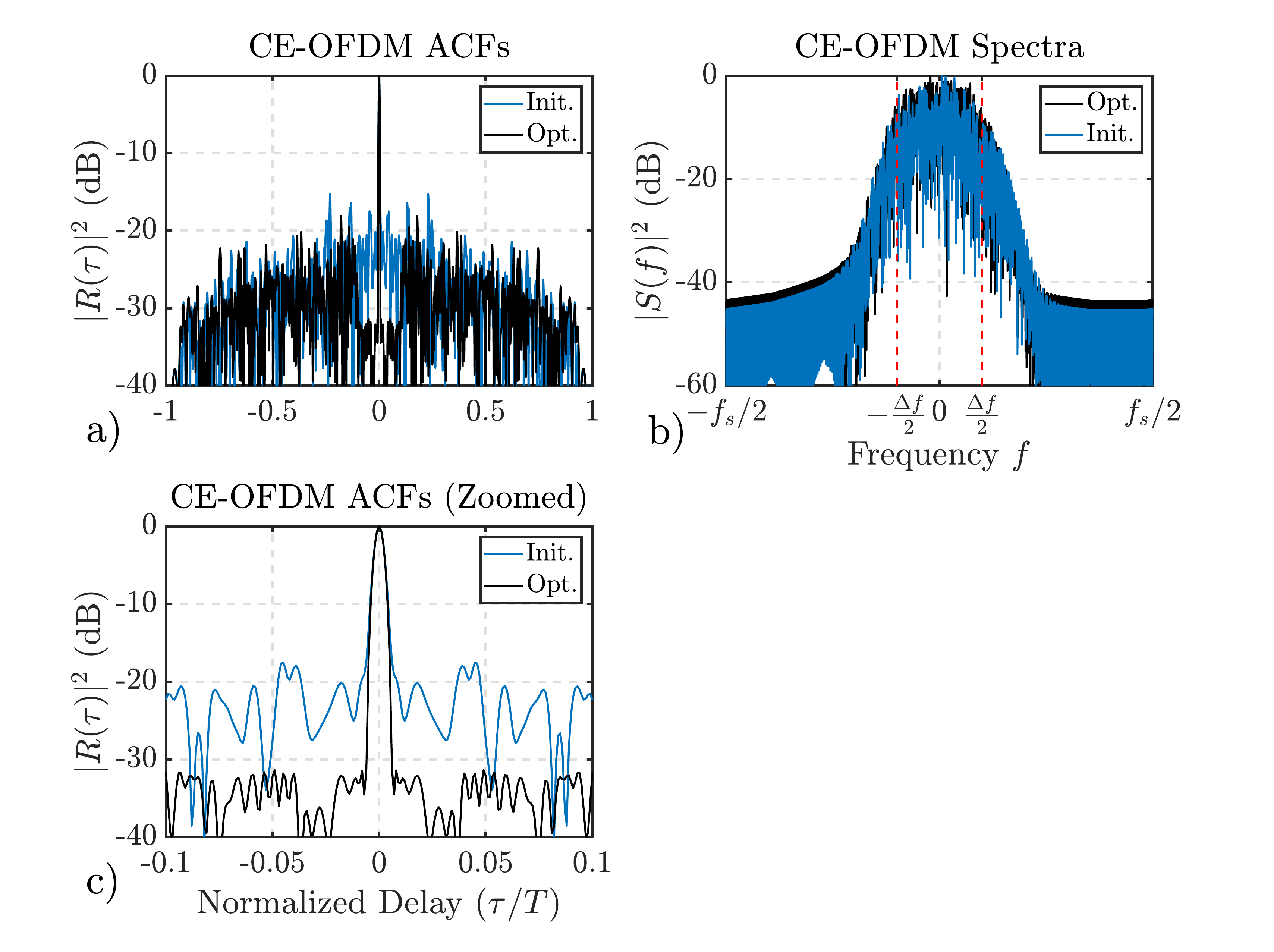}
\caption{ACFs (a), Spectra (b), and zoomed ACFs (c) of initial and optimized CE-OFDM waveforms.  The CE-OFDM waveform resulting from the GISL optimization algorithm possesses clearly lower ACF sidelobes in the specified sub-region of time-delays while also maintaining the same mainlobe width. }
\label{fig:CEOFDM_3}
\end{figure}

Figure \ref{fig:CEOFDM_3} shows the result of optimizing the GISL of the CE-OFDM waveform from Figure \ref{fig:CEOFDM_1} over a sub-region of time-delays $\Omega_{\tau} \in \Delta \tau \leq |\tau| \leq 0.1T$.  The initial waveform's GISL and PSLR over this sub-region of time-delays was -16.8 dB and -17.51 dB respectively.  The resulting optimized waveform's GISL and PSLR values were -30.34 dB and -31.39 dB respectively, a substantial reduction of sidelobes in the specified region $\Omega_{\tau}$.  Again, the optimized waveform's mainlobe width stays fixed as does the waveform's spectral extent.  

\begin{figure}[ht]
\centering
\includegraphics[width=0.5\textwidth]{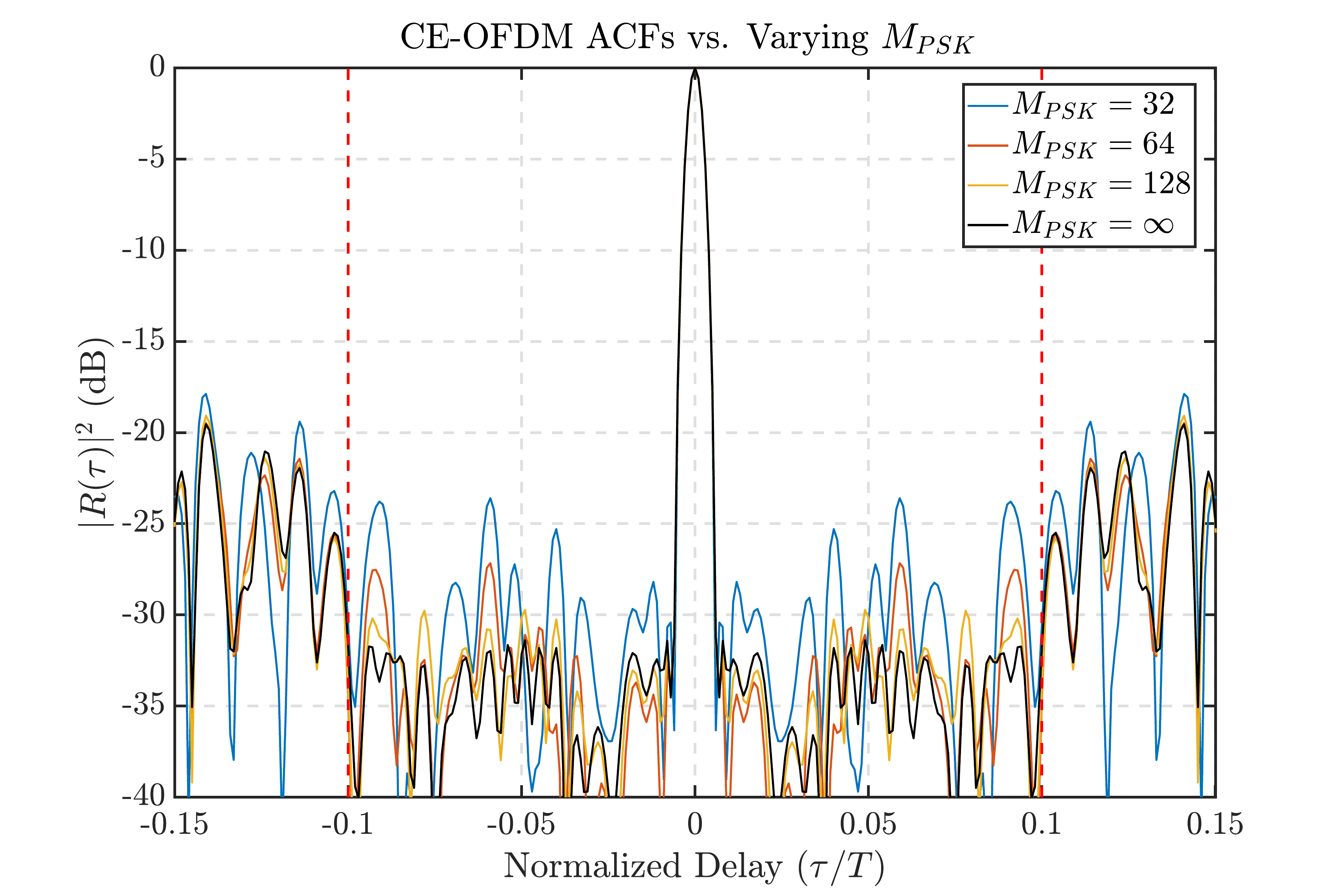}
\caption{ACF of the CE-OFDM in Figure \ref{fig:CEOFDM_3} and corresponding ACFs of the same waveform using finite M-ary alphabets to represent the PSK symbols.  Lower $M_{\text{PSK}}$ introduces stronger perturbations to the original waveform's PSK symbols.  This in turn degrades the original waveform's desirably low ACF sidelobe levels. }
\label{fig:CEOFDM_5}
\end{figure}

One final important point is that the GD-GISL optimization routine assumes the CE-OFDM waveforms employ PSK with infinite granularity (i.e, $M_{\text{PSK}}=\infty$).  Practical implementations of CE-OFDM waveforms will almost certainly employ a finite M-ary alphabet for the PSK symbols.  Truncating these continuous symbols to an M-ary representation will introduce perturbations in the waveform's instantaneous phase.  These perturbations are likely to degrade the desirably low ACF sidelobes of the $M_{\text{PSK}}=\infty$ optimized waveform designs.  This issue is illustrated in Figure \ref{fig:CEOFDM_5} where the PSK symbols from the optimized waveform in Figure \ref{fig:CEOFDM_3} is approximated by finite M-ary alphabets.  As can be seen from the figure, as $M_{\text{PSK}}$ is reduced, the degree of perturbations in the PSK symbols is greater and the ACF sidelobes increase in the region of time-delays $\Omega_{\tau}$ where the optimization routine was run on.  There are two ways to mitigate this issue.  The first is to utilize a large $M_{\text{PSK}}$ value to reduce the degree of perturbations in the implemented waveform.  The second is to modify the GD-GISL algorithm to operate on finite M-ary alphabets.  This approach may produce more favorable optimal designs with less perturbations than the waveforms shown in Figure \ref{fig:CEOFDM_5}.  This will be a topic of future investigation.

\section{Conclusion}
\label{sec:Conclusion}
The GD-GISL minimization algorithm synthesizes CE-OFDM waveforms with low ACF sidelobes in a specified region of time-delays.  The algorithm leverages methods developed in \cite{Blunt_PCFM_Grad} which  exploits FFTs in the majority of its operations making it computationally feasible to optimize CE-OFDM waveforms with a large number of sub-carriers $L$.  Results show that the algorithm is indeed capable of reducing CE-OFDM ACF sidelobes in a user-specified sub-region of time-delays while also maintaining the waveform's ACF mainlobe width.  Truncating these PSK symbols to M-ary alphabets introduces perturbations to the optimal design that results in increased ACF sidelobes.  There are a number of avenues to pursue for future work.  The first obvious one is to modify the algorithm to work with finite M-ary alphabets for PSK and other digital modulation techniques such as Quadrature Amplitude Modulation (QAM).  The algorithm should also be generalizable to generating families of waveforms with desirable ACF and Cross-Correlation Function (CCF) properties such as the efforts of \cite{PrabhuBabuI} and extendable to minimizing $\ell_p$-norms on the AF of the waveform as well.   Lastly, these optimization algorithms should also be applicable to the Multi-Tone Sinusoidal FM (MTSFM) waveform \cite{Hague_AES} of which the CE-OFDM waveform is a special case.   

\appendix
\section{Derivation of the Gradient of the GISL Metric}
\label{sec:gislMyNisl}
We define the cost metric as the GISL in \eqref{eq:GISLmyNISL} as
\begin{equation}
J_p = \dfrac{\|\mathbf{w}_{\text{SL}} \odot \mathbf{r} \|_p^2}{\|\mathbf{w}_{\text{ML}} \odot \mathbf{r} \|_p^2}=\left(\dfrac{\mathbf{w}_{\text{SL}}\trans  |\mathbf{r}|^p}{\mathbf{w}_{\text{ML}}\trans  |\mathbf{r}|^p}\right)^{2/p}.
\label{eq:A_1}
\end{equation}
The Wirtinger definition of a complex derivative extended to vector calculus is \cite{cookbook}
\begin{equation} \label{eq:wirtDef}
	\grad{\mathbf{x}}{}= \begin{cases}
			\left( \fracpar{\mathbf{y}}{\mathbf{x}}\right)\trans \grad{\mathbf{y}}{} + \left( \fracpar{\mathbf{y}^*}{\mathbf{x}}\right)\trans \grad{\mathbf{y^*}}{} & \mathbf{y}\in\mathbb{C} \\
			\left( \frac{\partial \mathbf{y}}{\partial \mathbf{x}}\right)\trans \grad{\mathbf{y}}{}   & \mathbf{y}\in\mathbb{R}
	\end{cases}
\end{equation}
where for an arbitrary $\left\{ \mathbf{y}(\mathbf{x}) : n_x \to n_y \right\} $ transformation from a $ n_x $ to a $ n_y $ dimensional space,
\begin{equation}
	\fracpar{\mathbf{y}}{\mathbf{x}} = \begin{bmatrix}
		\grad{\mathbf{x}}{y_1} & \grad{\mathbf{x}}{y_2} & \dots & \grad{\mathbf{x}}{y_{n_y}}
	\end{bmatrix}\trans
\end{equation} 
is the $ n_y {\times} n_x $ Jacobian matrix.

We begin with the real to complex mapping from $ \bphi $ to $ \mathbf{\bar{s}}^* $
\begin{equation} \label{eq:dphi}
	\begin{aligned}
		\grad{\bphi}{J_p} &= \left( \fracpar{\mathbf{\bar{s}^*}}{\bphi}\right)\trans \grad{\mathbf{\bar{s}^*}}{J_p} + \left( \fracpar{\mathbf{\bar{s}}}{\bphi}\right)\trans \grad{\mathbf{\bar{s}}}{J_p} \\
		&= -j\mathbf{\bar{D}}\trans \left(\mathbf{\bar{s}}^* \odot \grad{\mathbf{\bar{s}^*}}{J_p} \right) + j\mathbf{\bar{D}}\trans \left(\mathbf{\bar{s}} \odot \grad{\mathbf{\bar{s}}}{J_p} \right)\\
		&= 2 \mathbf{\bar{D}}^T \Im{\left\{\mathbf{\bar{s}}^* \odot \grad{\mathbf{\bar{s}^*}}{J_p} \right\}} 
	\end{aligned}
\end{equation} 
where the matrix
\begin{equation}
	\mathbf{\bar{D}}=-\mathbf{\bar{B}}_c \text{diag}\left\{\sin\left(\bphi\right)\right\}+\mathbf{\bar{B}}_s \text{diag}\left\{\cos\left(\bphi\right)\right\}
\end{equation}
results from the chain rule, and where $ \text{diag} \left\{\bullet\right\} $ places the elements of the operand along the diagonal of a square matrix such that the $ \ell\nth $ column is 
\begin{equation}
	\mathbf{\bar{d}}_\ell=-\mathbf{\bar{b}}_{c,\ell} \sin\left(\phi_\ell\right)+\mathbf{\bar{b}}_{s,\ell} \cos\left(\phi_\ell\right)
\end{equation}
Note that $ \mathbf{\bar{B}}_c $, $\mathbf{\bar{B}}_s $, and $ \mathbf{\bar{D}} $ are zero padded such that their dimensionality is now $( 2M-1){\times}L $. Next, we map from $ \mathbf{\bar{s}}^* $ to $ (\mathbf{A\bar{s}})^* $
\begin{equation} \label{eq:dsconj}
	\begin{aligned}
		\nabla_{ \mathbf{\bar{s}^*}} J_p &= \left( \fracpar{(\mathbf{A\bar{s}})^* }{\mathbf{\bar{s}}^*}\right)\trans\grad{(\mathbf{A\bar{s}})^*}{J_p}+ \left( \fracpar{(\mathbf{A\bar{s}})}{\mathbf{\bar{s}}^*}\right)\trans \grad{(\mathbf{A\bar{s}})}{J_p} \\
		&= \mathbf{A}\herm \grad{(\mathbf{A\bar{s}})^*}{J_p}
	\end{aligned}
\end{equation}
where the second term simplifies to 0 since $  \mathbf{\bar{s}} $ and $ \mathbf{\bar{s}}^* $ are considered independent variables. Next, mapping from $ (\mathbf{A\bar{s}})^* $ to $ \lvert\mathbf{A\bar{s}}\rvert^2 $ yields 
\begin{equation} \label{eq:dAsconj}
	\begin{aligned}
		\grad{(\mathbf{A\bar{s}})^*}{J_p} &= \left( \fracpar{\lvert\mathbf{A\bar{s}}\rvert^2}{(\mathbf{A\bar{s}})^*}\right)\trans\grad{\lvert\mathbf{A\bar{s}}\rvert^2}{J_p} \\
		&= \left( \fracpar{\left(\mathbf{A\bar{s}}\odot \left(\mathbf{A\bar{s}}\right)^*\right)}{(\mathbf{A\bar{s}})^*}\right)\trans\grad{\lvert\mathbf{A\bar{s}}\rvert^2}{J_p} \\
		&= \text{diag}\left\{\mathbf{A\bar{s}}\right\} \grad{\lvert\mathbf{A\bar{s}}\rvert^2}{J_p} \\
		&= (\mathbf{A\bar{s}}) \odot \grad{\lvert\mathbf{A\bar{s}}\rvert^2}{J_p}
	\end{aligned}
\end{equation}
Then, since $ \mathbf{r} $ is a function of $ \lvert\mathbf{A\bar{s}}\rvert^2 $ via \eqref{eq:discr}, 
\begin{equation} \label{eq:dAs2}
	\begin{aligned}
		\grad{\lvert\mathbf{A\bar{s}}\rvert^2}{J_p}&=\left( \fracpar{\mathbf{r}}{\lvert\mathbf{A\bar{s}}\rvert^2}\right)\trans\grad{\mathbf{r}}{J_p} + \left( \fracpar{\mathbf{r}^*}{\lvert\mathbf{A\bar{s}}\rvert^2}\right)\trans\grad{\mathbf{r^*}}{J_p} \\
		&= \mathbf{A}^*  \grad{\mathbf{r}}{J_p} + \mathbf{A}  \grad{\mathbf{r^*}}{J_p} \\
		&= 2 \Re\left\{\mathbf{A} \grad{\mathbf{r^*}}{J_p}\right\}
	\end{aligned}
\end{equation}
And, mapping from $ \mathbf{r^*} $ to $ \lvert\mathbf{r}\rvert^p $ 
\begin{equation} \label{eq:drconj}
	\begin{aligned}
			\grad{\mathbf{r^*}}{J_p} &= \left(\fracpar{\lvert\mathbf{r}\rvert^p}{\mathbf{r^*}}\right)\trans\grad{\lvert\mathbf{r}\rvert^p}{J_p} \\
			&= \left(\fracpar{\left(\mathbf{r}\odot\mathbf{r^*}\right)^{p/2}}{\mathbf{r^*}}\right)\trans\grad{\lvert\mathbf{r}\rvert^p}{J_p} \\
			&=   \frac{p}{2} \text{diag}\left\{\lvert\mathbf{r}\rvert^{p-2} \odot \mathbf{r}\right\}\grad{\lvert\mathbf{r}\rvert^p}{J_p} \\
			&= \frac{p}{2} \lvert\mathbf{r}\rvert^{p-2}\odot\mathbf{r}\odot \grad{\lvert\mathbf{r}\rvert^p}{J_p}
	\end{aligned}
\end{equation} 
And, since $ J_p $ is directly a function of $ \lvert\mathbf{r}\rvert^p $, we solve for $ \grad{\lvert\mathbf{r}\rvert^p}{J_p} $ by use of the quotient rule followed by chain rule

\begin{align} \label{eq:drp}
		 \grad{\lvert\mathbf{r}\rvert^p}{J_p} &= \frac{2}{p} \left(\frac{\wSL\trans \vert \mathbf{r} \vert^p}{\wML\trans \vert \mathbf{r} \vert^p}\right)^{2/p-1} \IEEEnonumber \\ &  \frac{\left(\den\right)\fracpar{\left(\num\right)}{\lvert\mathbf{r}\rvert^p} - \left(\num\right)\fracpar{\left(\den\right)}{\lvert\mathbf{r}\rvert^p}}{\left(\den\right)^2} \IEEEnonumber \\
		&=\frac{2}{p} J_p \left(\frac{\den}{\num}\right)  \frac{\left(\den\right)\wSL - \left(\num\right)\wML}{\left(\den\right)^2} \IEEEnonumber\\
		&= \frac{2}{p} J_p \left[ \frac{\wSL}{\num} - \frac{\wML}{\den} \right]
\end{align}

Finally, using \eqref{eq:dAsconj}-\eqref{eq:drp} and inserting \eqref{eq:dsconj} into \eqref{eq:dphi} produces the final result
\begin{align}
	\grad{\bphi}{J_p} &= 2 \mathbf{\bar{D}}\trans \Im{\left\{\mathbf{\bar{s}}^* \odot \grad{\mathbf{\bar{s}^*}}{J_p} \right\}} \IEEEnonumber \\
	&= 2 \mathbf{\bar{D}}\trans \Im{\left\{\mathbf{\bar{s}}^* \odot \mathbf{A}\herm \grad{(\mathbf{A\bar{s}})^*}{J_p} \right\}} \IEEEnonumber \\
	&= 2 \mathbf{\bar{D}}\trans \Im{\left\{\mathbf{\bar{s}}^* \odot \mathbf{A}\herm \left[(\mathbf{A\bar{s}}) \odot \grad{\lvert\mathbf{A\bar{s}}\rvert^2}{J_p}\right] \right\}}\IEEEnonumber \\
	&= 2 \mathbf{\bar{D}}\trans \Im{\left\{\mathbf{\bar{s}}^* \odot \mathbf{A}\herm \left[(\mathbf{A\bar{s}}) \odot 2 \Re\left\{\mathbf{A} \grad{\mathbf{r^*}}{J_p}\right\}\right] \right\}}\IEEEnonumber \\
	&= 4 J_p \mathbf{\bar{D}}\trans \Im{\left\{\mathbf{\bar{s}}^* \odot \mathbf{A}\herm \left[(\mathbf{A\bar{s}}) \odot   \mathbf{P} \right]\right\}} 
\label{eq:resultRe}
	\end{align}
where 
\begin{equation}
\mathbf{P} = \Re\left\{\mathbf{A} \left( \lvert\mathbf{r}\rvert^{p-2}\odot\mathbf{r}\odot \left[ \frac{\wSL}{\num} - \frac{\wML}{\den} \right]\right)\right\}.
\end{equation}
Note that a function is said to be conjugate symmetric if and only if its Fourier transform is real valued \cite{Proakis}. Under the assumption that $ \wSL $ and $ \wML $ are defined to be symmetric about 0 delay, then $ \left( \lvert\mathbf{r}\rvert^{p-2}\odot\mathbf{r}\odot  \left[ \frac{\wSL}{\num} - \frac{\wML}{\den} \right]\right) $ term exhibits conjugate symmetry since by definition, a waveform's ACF is conjugate symmetric.  Additionally, since we are applying a DFT matrix to a conjugate symmetric vector, the result is purely real and so the $ \Re\{\bullet\} $ operator can be discarded.

\end{document}